\newcommand{\balpha}{\mbox{\boldmath$\alpha$}}

\newcommand{\btheta}{\mbox{\boldmath$\theta$}}

\newcommand{\bx}{\mbox{\boldmath$x$}}

\documentclass[10pt]{article}
\usepackage[top=1in, left=1in, right=1in, bottom=1in]{geometry}
\usepackage{amsmath,amssymb,amsthm,graphicx,algorithm,algorithmic}
\title{Approximate Maximum A Posteriori Inference with Entropic Priors}
\author{Matthew D. Hoffman \\
Department of Statistics \\
Columbia University \\
New York, NY \\
mdhoffma@cs.princeton.edu}
\date{}

\begin{document}

\newcommand{\bm}[1]{\mbox{\boldmath{$#1$}}}

\maketitle

\begin{abstract}
  In certain applications it is useful to fit multinomial
  distributions to observed data with a penalty term that encourages
  sparsity. For example, in probabilistic latent audio source
  decomposition one may wish to encode the assumption that only a few
  latent sources are active at any given time. The standard heuristic
  of applying an L1 penalty is not an option when fitting the
  parameters to a multinomial distribution, which are constrained to
  sum to 1.  An alternative is to use a penalty term that encourages
  low-entropy solutions, which corresponds to maximum a posteriori
  (MAP) parameter estimation with an entropic prior. The lack of
  conjugacy between the entropic prior and the multinomial
  distribution complicates this approach. In this report I propose a
  simple iterative algorithm for MAP estimation of multinomial
  distributions with sparsity-inducing entropic priors.
\end{abstract}

\section{Introduction}
Suppose we want to estimate the parameter $\btheta$ to a multinomial
distribution responsible for generating $N$ observations
$x_i\in\{1,\ldots,K\}$.  The log-likelihood of the data is given by
\begin{equation}
\log p(\bx) = \sum_i \log \theta_{x_i},
\end{equation}
and the maximum-likelihood estimate of $\btheta$ is simply
\begin{equation}
\theta_k \propto \sum_i \mathbb{I}[x_i = k],
\end{equation}
where $\mathbb{I}$ is an indicator function whose value is 1 if its
argument is true and 0 if its argument is false.

The maximum-likelihood estimate may not be optimal if we have a priori
knowledge that leads us to believe that $\btheta$ is sparse. For
example, if $\btheta$ indicated the relative loudness of a set of 88
piano notes at particular moment (as it might in an application of
Probabilistic Latent Component Analysis to audio spectrograms
\cite{Smaragdis:2006}), then we might expect only a few elements of
$\btheta$ to be much greater than 0. This would correspond to the
intuition that pianists rarely mash the entire piano keyboard at
once.

To incorporate this prior intuition into our analysis, we might
add a penalty term to our log-likelihood function that encourages
sparse settings of $\btheta$. A common heuristic for inducing sparsity
in optimization problems is to introduce an L1 penalty term into the
cost function (for example, in lasso regression \cite{Hastie:2001}).
This is not an option here, since the L1 norm of $\btheta$ is 
constrained to be 1. A natural alternative is to include a negative
entropy term in the log-likelihood function, corresponding to placing
an unnormalized sparse entropic prior on $\btheta$:
\begin{equation}
\label{eq:joint}
\log p(\btheta, \bx) = \textrm{constant} + a \sum_k \theta_k\log \theta_k
+ \sum_i \log \theta_{x_i}.
\end{equation}
The constant $a$ controls the strength of the prior 
$p(\btheta)\propto\exp\{a\sum_k \theta_k\log \theta_k\}$. If $a$ is
positive, then this prior will give higher weight to low-entropy
settings of $\btheta$.

Unfortunately, the Maximum A Posteriori (MAP) estimate of $\btheta$
does not have a simple analytic form for this model, since the
entropic prior is not conjugate to the multinomial distribution. In
the following section, I propose a simple iterative scheme for MAP
estimation of $\btheta$ when $a$ is positive.

\section{A MAP Inference Scheme for the Sparse Entropic Prior}

Our strategy is based on optimizing the following approximate
auxiliary function for the negative entropy term in equation
\ref{eq:joint}:
\begin{equation}
\ell(a, \nu, \btheta, \balpha) \triangleq
a\sum_k \alpha_k (\nu \log\theta_k - (\nu-1)\log\alpha_k),
\end{equation}
where $\balpha$ is a free parameter such that $\sum_k\alpha_k=1$ and
$\alpha_k\ge 0$, and $\nu$ is a real-valued scalar constrained to be
greater than 1. Taking the derivative of the Lagrangian of $\ell$
with respect to $\alpha_k$ yields
\begin{equation}
\frac{\partial \ell}{\partial \alpha_k}
= a\nu\log\theta_k - a(\nu-1)(1 + \log\alpha_k) + \lambda.
\end{equation}
Setting the right side equal to zero shows that $\ell$ is optimized
with respect to $\balpha$ when
\begin{equation}
\alpha_k \propto \exp\left\{\frac{\nu}{\nu-1}\log\theta_k\right\}
= \theta_k^{\frac{\nu}{\nu-1}}.
\end{equation}
When $\nu$ is large, this implies that the optimal value of $\ell$
can only be achieved when $\alpha_k \approx \theta_k$.

When $\alpha_k=\theta_k$, we recover the original entropic prior
term:
\begin{equation}
\ell(a, \nu, \btheta, \btheta)
= a\sum_k \theta_k (\nu\log\theta_k - (\nu-1)\log\theta_k)
= a\sum_k \theta_k\log\theta_k.
\end{equation}
Thus, for sufficiently large values of $\nu$, when $\balpha$ is
optimally chosen $\ell$ approximates the entropic prior. We may
therefore substitute $\ell$ (with a large value of $\nu$) for
the entropic prior term in equation \ref{eq:joint} and jointly
optimize the approximate objective
\begin{equation}
\label{eq:approxjoint}
\mathcal{L} \triangleq 
a\sum_k \alpha_k (\nu \log\theta_k - (\nu-1)\log\alpha_k)
+ \sum_i \log \theta_{x_i}.
\end{equation}
over $\balpha$ and $\btheta$. When the gradient of $\mathcal{L}$ with
respect to $\balpha$ is 0, as it must be at a local optimum of
$\mathcal{L}$, $\alpha_k \approx \theta_k$, and so $\mathcal{L}
\approx \log p(\bx, \btheta)$, the objective function of interest.

A simple fixed-point iteration can be used to optimize $\mathcal{L}$
over $\balpha$ and $\btheta$. The gradient of the Lagrangian with
respect to $\theta_k$ is
\begin{equation}
\frac{\partial \mathcal{L}}{\partial \theta_k}
= \frac{1}{\theta_k}\left(a\alpha_k\nu + \sum_i \mathbb{I}[x_i=k]\right)
+ \lambda,
\end{equation}
where $\mathbb{I}$ is an indicator function whose value is 1 if its
argument is true and 0 if its argument is false. $\mathcal{L}$ is
therefore maximized with respect to $\btheta$ when
\begin{equation}
\label{eq:thetaupdate}
\theta_k \propto a\alpha_k\nu + \sum_i \mathbb{I}[x_i=k].
\end{equation}
As observed above, $\mathcal{L}$ is maximized with respect to $\balpha$
when
\begin{equation}
\label{eq:alphaupdate}
\alpha_k \propto \theta_k^{\frac{\nu}{\nu-1}}.
\end{equation}
By iterating between the updates in equation \ref{eq:thetaupdate} and
\ref{eq:alphaupdate}, we reach a stationary point of $\mathcal{L}$.
At such a stationary point, $\mathcal{L}\approx\log p(\bx, \btheta)$,
and so we may conclude that the value of $\btheta$ at a stationary
point of $\mathcal{L}$ yields approximately a local optimum of $\log
p(\bx, \btheta)$.

Note that these updates may require a number of iterations to
converge, and the number of iterations needed is likely to grow with
$\nu$. However, the cost of each update is minimal. If these updates
are incorporated as part of a larger coordinate ascent algorithm like
the expectation-maximization algorithm used in probabilistic latent
semantic indexing \cite{Hofmann:1999b}, the additional expense
involved in iterating between updating $\btheta$ and $\balpha$ is
likely to be dominated by the cost of computing $\sum_i
\mathbb{I}[x_i=k]$ (or its expected value).

\section{Conclusion}
I have presented a simple fixed-point iteration for performing
approximate maximum a posteriori estimation of multinomial parameters
in the presence of a sparsity-inducing entropic prior. This algorithm
only provides an approximate solution, but it can be made arbitrarily
accurate at the cost of slower convergence. The algorithm is very
easy to implement, and the cost per iteration is minimal.

\bibliographystyle{unsrt}
\bibliography{mdh-entropic}

\end{document}